\documentclass{elsart} 
\usepackage{amssymb}
\usepackage{amsmath}
\usepackage{graphicx}
\usepackage{graphics}
\usepackage{epsf}
\begin{document}
\newcommand{\eq}{\begin{equation}}                                
\newcommand{\eqe}{\end{equation}}  

\begin{frontmatter}

\title{Searching for dark matter with helium atom}

\author{I. F. Barna}
\ead{barnai@sunserv.kfki.hu}
\address{ Central Physical Research Institute (KFKI),  Radiation and Environmental Physics Department in the  
Atomic Energy Research Institute, P.O. Box 49, H-1525 Budapest, Hungary, EU }    

\begin{abstract}

With the help of the boost operator we can model the interaction between a  
weakly interacting particle(WIMP) of dark matter(DAMA) and an atomic nuclei. 
Via this ``kick" we calculate the total electronic excitation cross section of the helium atom. 
The bound spectrum of He is calculated through a diagonalization process with a 
configuration interaction (CI) wavefunction built up from Slater orbitals. All together 
19 singly- and doubly-excited atomic sates were taken with total angular momenta of 
L=0,1 and 2. Our calculation may give a rude estimation about the magnitude of the total excitation cross 
section which could be measured in later scintillator experiments. The upper limit of the
excitation cross section is $9.7\cdot 10^{-8}$ barn. 
   
\end{abstract}
\end{frontmatter}

\section{Introduction}                                    
\label{intro}                                    
Searching for DAMA with WIMP is an interesting question from both theoretical \cite{jung}
and experimental sides. 
A considerable experimental work are in progress to measure DAMA-nucleus interaction 
in different scintillation set-ups such as liquid xenon \cite{bern}, 
NaI \cite{gerb} or different anisotropic crystals \cite{bern2,belli}. More technical details 
about the running experiments can be found under \cite{web}.
Theoretical considerations state that the WIMP can scatter from 
a nucleus either via a scalar (spin-independent) interaction or via an axial-vector 
(spin-dependent) interaction \cite{pie}. 

In the following we use the binary encounter approach and apply the boost operator, to 
model the ``kick" between the unknown WIMP-DAMA and the nucleus of the helium atom 
and calculate the total electronic excitation cross section. 
We use a CI wavefunction built up from Slater-like orbitals to 
describe the ground state and the low-lying bound spectrum of helium.
Our CI wavefunction was successfully applied to describe different time-dependent
problems such as heavy-ion helium collisions \cite{bar1,bar2} or photoionization of helium in 
short XUV laser fields \cite{bar3}. 
According to our knowledge, there are no theoretical calculations from 
this type forecasting measurable total excitation cross sections. \\
 Atomic units are used throughout the paper unless otherwise mentioned.

\section{Theory}

At first we have to calculate the low-lying bound spectrum of the He. 
We obtain the eigenfunctions and the eigenvalues by 
diagonalizing the time-independent Schr\"odinger equation 
\eq
\hat{H}_{He} \Phi_j = E_j \Phi_j,
\eqe
where $\hat{H}_{He}$ is the spin independent Hamiltonian of the unperturbed helium atom
\eq
\hat{H}_{He} = \frac{{\bf{p}}_1^2}{2} + \frac{{\bf{p}}_1^2}{2}- 
\frac{2}{{\bf{r}}_1} - \frac{2}{{\bf{r}}_2}
+ \frac{1}{|{\bf{r}}_1- {\bf{r}}_2|},
\eqe
and $\Phi_j$ is the CI wavefunction built up by a finite linear 
combination of symmetrized products 
of Slater orbitals
\eq
\phi({\bf{r}}_1) = c(n,\kappa)r^{n-1} e^{-r \kappa} Y_{l,m}(\theta,\varphi),
\eqe
where $c(n,\kappa)$ is the normalization constant. 
We use Slater functions with angular momentum $l=0,1,2$ and couple 
them to $L=0,1$ and 2 total angular momentum two-electron states. 
In our basis we apply 9 different s orbitals, 6 different p orbitals and 4 different d 
orbitals, respectively.  
Table I presents  our bound He spectrum compared to other, much larger {\it{ab initio}} 
calculations \cite{burg,has,ho}. We implement the complex scaling \cite{nim} method to 
identify the double-excited states in the low-lying single continuum.
It is well known that the 1s1s ground state is highly angular correlated, and further  $pp$ and 
$dd$ terms are needed to have a accurate agreement with experimental data which is 
$-2.904$ a.u. \cite{bar3}. We checked the role of these terms and found that the affect in the final total 
cross sections is negligible.

We may approximate the interaction between the unknown DAMA particle and the 
nucleus of the helium with 
the boost operator. If we suddenly ``kick" the He nucleus with a ${\bf{k}}$ boost  
in the direction of ${\bf{r}}$ that is equivalent to a collective ${-\bf{k}}/2$
``kick" of the two atomic electrons according to the center-of-mass. For a better understanding 
the geometry of the interaction is presented in Figure 1. 

The total excitation amplitude can be calculated in the following way:
\eq
a_{exc} = \sum_{f} \langle \Phi({\bf{r}}_1,{\bf{r}}_2)_{f} | 
e^{-i{\bf{r}_1} {\bf{k}}/2 - i{\bf{r}_2} {\bf{k}}/2}  | 
\Phi({\bf{r}}_1,{\bf{r}}_2)_{1s1s} \rangle, 
\eqe
where $1s1s$ is the ground state of He 
and the summation $f$  runs over the singly- and doubly-excited final states. 
For elastic collision only the ground state to ground state transition 
is considered.  
The energy of the unknown DAMA is $E = k^2/2$. 
The DAMA-electron interaction is evaluated thought the
 transition matrix elements of the boost 
operator between two Slater orbitals
$ \langle \phi_1({\bf{r}} | e^{-i{\bf{r}} {\bf{k}}/2}  | 
\phi_2({\bf{r}}) \rangle $.
To separate the radial and the angular part of the matrix element we expand 
the plane wave through spherical Bessel functions in the well-known way
\cite{Mess}: 
\eq
e^{i{\bf{r}} \cdot {\bf{k}}} = e^{k \cdot r cos(\theta)} =
4\pi \sum_{l=0}^{\infty} \sum_{m=-l}^{+l} i^l j_l(kr) 
Y^*_{l,m}(\theta_{\bf{k}},\varphi_{\bf{k}})
 Y_{l,m}(\theta_{\bf{r}},\varphi_{\bf{r}}).
\eqe
After some algebra the angular part of the matrix element
 gives us the Clebsch-Gordan coefficient
 \begin{center} 
 \begin{eqnarray}
\int\limits_{\Omega} Y^*_{l_1,m_1}(\theta,\varphi)  Y_{l,m}(\theta,
\varphi)Y_{l_2,m_2}(\theta,\varphi) d \Omega =
{\sqrt{\frac{(2l_2+1)(2l+1)}{2\pi(2l_1+1)}}} \times \nonumber \\ 
(l_2,l,l_1|m_2,m,m_1) \cdot (l_2,l,l_1|0,0,0).
\end{eqnarray}
\end{center}
According to the definition of the spherical Bessel functions \cite{numrec}a
$j_l(kr) = \sqrt{\frac{\pi}{2kr}}J_{l+1/2}(kr)$ the radial part of the matrix element 
has the analytic solution of \cite{rizsik}:
 \begin{center} 
 \begin{eqnarray}
\int\limits_0^\infty J_{\nu}(k r )e^{-\alpha r}r^{\mu -1} dr = 
 \frac{ \left(\frac{k}{2}\right)^{\nu} \Gamma(\nu + \mu) }{ \Gamma(\nu + 1) 
 \sqrt{(k^2+\alpha^2)^{\nu +\mu}} }  \times  \nonumber  \\
   { _{2}\!F_1} \left( \frac{\nu + \mu}{2},\frac{1-\mu+\nu}{2} ; \nu + 1 ;
 \frac{k^2}{k^2+\alpha^2} \right),
 \end{eqnarray}
\end{center}
where $\Gamma$ is the gamma function and ${ _{2}\!F_1}$ is the hypergeometric function with 
the following real arguments $\mu-1 = n_1+n_2-1/2$, $\nu = l+ 1/2 $ and $\alpha = \kappa_1 +\kappa_2 $,  
We tried to simplify the final formula but unfortunately, we could not 
succeed, and have to calculate the hypergeometric function 
numerically with a well-behaving complex contour integral \cite{numrec}b.
It is worth to mention that with additional constraints among the parameters $[\alpha,k,\nu,\mu]$
 this radial integral can be simplified, but not in our general case.
 
The total excitation cross section can be evaluated with the following formula
\eq
\sigma_{exc} = r_{\alpha}^2 \pi P_{exc},
\eqe
where $r_{\alpha} = 1.76\cdot 10^{-15} m$ is the radius of the He nucleus and 
$P_{exc} =|a_{exc}|^2 $ is the total excitation probability.
For elastic collision only the ground state to ground state transition is considered. 

\section{Results and discussion}  
Figure 2 presents our elastic and excitation total cross sections  
in function of the impulse of the DAMA. 
The cross sections are given in barns and the wave number of the unknown particle 
is given in atomic units. If the velocity of the DAMA is known then the 
mass can be calculated from $m= k/v$.
 
The maximum of the excitation cross section is $9.7\times 10^{-8}$ barn 
at k=3 a.u. DAMA impulse. 
At low wave numbers (k) the boost operator can be well approximated with its 
Taylor series which is similar to the dipole interaction, and used in 
photoionization calculations.
At low k values the elastic cross section is many magnitude higher than the 
excitation one, which meets our physical intuition.   
At larger wave numbers, however, the dipole approximation breaks down and the general 
matrix element have to be calculated where the non-dipole contributions play 
a significant role. 
Above $k=30$ a.u. the elastic and excitation cross sections run together,  
but the excitation cross sections are a factor of 2-4 higher than the elastic ones.  
At wave numbers larger than 10, due to the quick oscillations of the boost operator,  
the cross sections have a strong decay which can be excellently fitted with the 
following power law:
\eq
\sigma_{exc} = 2.3566\times10^{5} \cdot k^{-13.782},
\eqe
 where the standard error of the exponent is 0.074 and the standard error 
 of the scaling constant is 0.48, respectively. 
 
At $k>10000$ wave numbers the cross sections stop decaying and show spurious oscillations,
which are numerical art-effects due to the limited accuracy of the calculations. These cross 
sections are not presented in Fig. 2. 

We can not enhance the total angular momenta of the two electron wavefunction in our 
calculation, and the number of the available bound states are also limited. 
The role of the highly-excited Rydberg states are out of our scope too.
In this sense we can not rigorously prove the convergence of our calculation, but 
our experience shows that for excitation the significant contributions always 
came from the lowest excited states.  
We interpret our results as a rude approximation for the dark matter He interaction 
which may stimulate further investigations.  
\section{Summary and Outlook}
With the help of the boost operator we gave a ``simple-man's model" for 
the DAMA-helium nucleus interaction   
and calculated total electronic excitation and elastic collision cross sections 
which can be measured in future scintillator experiments.  
Our calculation could be generalized for atoms with electrons more than two, 
(even for Xe) if the wavefunction of the ground state and a significant large 
number of excited states are present with sufficient accuracy.
We think that this problem could be solved with the General 
Relativistic Atomic Structure(GRASP) code \cite{gr}, which is out of our capability.  
The aim of this paper was twofold. First, we presented our model for
 the DAMA-He interaction calculating cross sections. Secondly, we  
advised our model to many-electron-atom theorists to calculate DAMA-Xe interaction. 
\section{Acknowledgment}
We thank Prof. J. Burgd\"orfer (TU Vienna) fur useful discussions and comments.
This work was not supported by any military agencies.

\newpage

\begin{figure}
\scalebox{0.45}{\rotatebox{-90}{\includegraphics{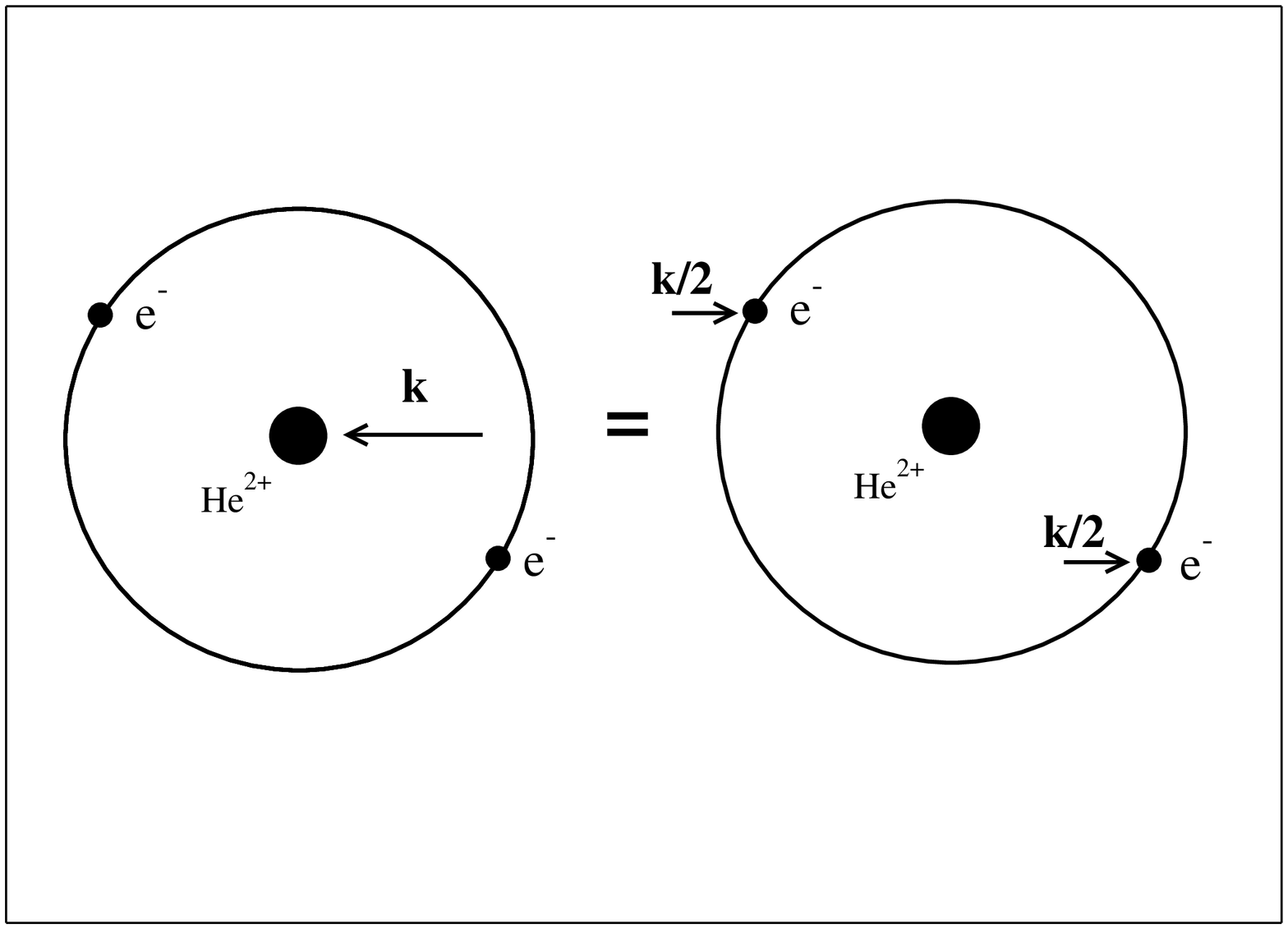}}}  
\caption{The geometry of the DAMA-He interaction.}
\end{figure}

\begin{table}
\caption{The energy levels of bound, singly-  and doubly-excited 
 states used in our calculations, compared with 
a) basis set calculations from \cite{burg} 
  b)  CI calculation results \cite{has} 
 and c) complex-coordinate rotation calculations from \cite{ho}. }              
\begin{center}                                                                  
 \begin{tabular}{lcc|lcc|lcc} \hline \hline
  states &    our  & other  & states & our      & other 
   & states & our &other \\
        &  results & theory  &       &   results & theory  &  &  results & theory    \\ \hline              
  1s1s & $ -2.8821 $ & $ -2.9037^a $ &  1s2p & $-2.1233 $ & $ -2.1238^b$  &   1s3d & $ -2.0556 $ & $-2.0556^b $   \\ 
  1s2s & $ -2.1441 $&$ -2.1460^a $&  1s3p &$ -2.0550 $ & $ -2.0551^b$     &  1s4d & $ -2.0312 $ & $-2.0313^b $ \\
  1s3s & $ -2.0607 $&$ -2.0612^a $&  1s4p &$ -2.0259 $ & $ -2.0310^b$ &  2s3d & $ -0.5597 $ & $-0.5692^c $ \\
  1s4s & $ -2.0333 $&$ -2.0335^a $&  2s2p &$ -0.6572 $ & $ -0.6931^c$ & 2s4d & $ -0.5305 $ & $-0.5564^c $ \\
  2s2s & $ -0.7297 $&$ -0.7779^a $&  2s3p &$ -0.5821 $ & $ -0.5971^c$ \\ 
  2s3s & $ -0.5711 $&$ -0.5899^a $&  2s4p &$ -0.5401 $ & $ -0.5640^c$  \\ 
  2s4s & $ -0.5372 $&$ -0.5449^a $&  2s5p &$ -0.5225 $ & $ -0.5470^c$ \\ 
  2s5s & $ -0.5133 $&$ -0.5267^a $&  3s3p &$ -0.2998 $ & $ -0.3356^c$  \\
   \hline \hline
  \end{tabular}
  \end{center}
  \end{table}
  

\begin{figure}
\scalebox{0.45}{\rotatebox{-90}{\includegraphics{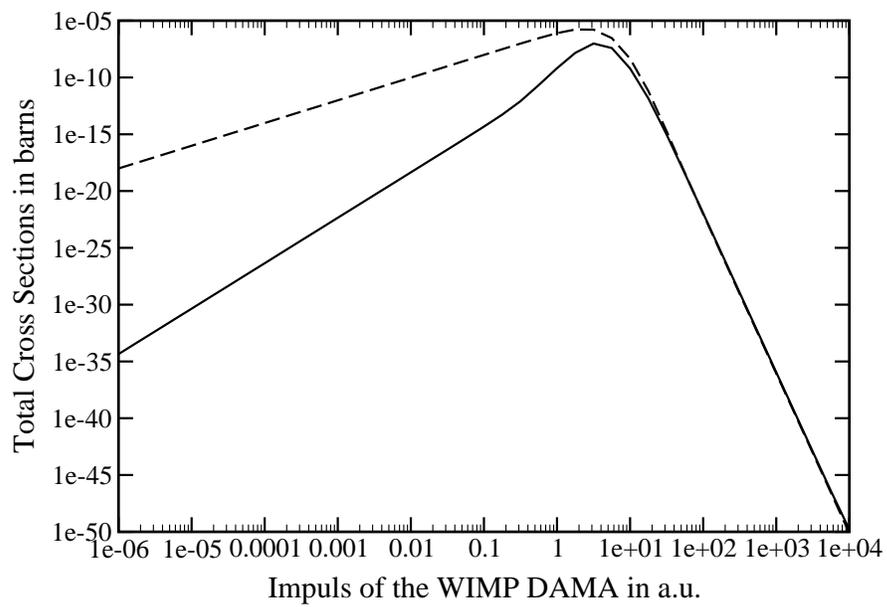}}}  
\caption{Elastic (dashed line) and excitation (solid line) total cross 
sections for DAMA-He collision.}
\end{figure}

\end{document}